\documentclass[traditabstract,referee]{aa}
\usepackage{graphicx}
\usepackage{setspace}
\usepackage{longtable,lscape}
\usepackage{hhline}
\usepackage{array}
\usepackage{natbib}
\usepackage{txfonts}
\title{\textbf{Finding proto-spectroscopic binaries}}
\subtitle{Precise multi-epoch radial velocities of 7 protostars in $\rho$ Ophiuchus
\thanks{Based on observations collected with the CRIRES spectrograph at the 
VLT/UT1 8.2-m Antu Telescope (ESO run ID. 081.C-0395(A)) at the Paranal Observatory, Chile}}%\footnotemark}
\author{P. Viana Almeida
          \inst{1,2,3}
	\and
          C. Melo
	  \inst{2}	
	\and
          N.C. Santos
	  \inst{1,3}
	\and
          P. Figueira
	  \inst{1}
	\and
	  M. Sterzik
		\inst{2}
	\and
	  J.F. Gameiro
		\inst{1,3}
          }
\institute{Centro de Astrof{\'i}sica, Universidade do Porto, Rua das Estrelas, 4150-762 Porto, Portugal
         \and
             ESO, Alonso de Cordova 3107, Casilla 19001, Vitacura, Santiago, Chile\\
             \email{palmeida@eso.org}
	  \and
	     Departamento F\'{\i}sica e Astronomia, Faculdade de Ci\^{e}ncias da Universidade do Porto, Portugal
		   }
  \abstract
  {The formation of spectroscopic binaries (SB) may be a natural byproduct of star formation. 
The early dynamical evolution of multiple stellar systems after the initial 
fragmentation of molecular clouds leaves characteristic imprints on the properties of young, multiple stars.
The discovery and the characterization of the youngest SB will allow us 
to infer the mechanisms and timescales involved  in their formation.
Our work aims to find spectroscopic companions around young stellar objects (YSO). 
We present a near-IR high-resolution (R~$\sim$~60000) 
multi-epoch radial velocity survey of 7 YSO in the star forming region (SFR) $\rho$ Ophiuchus. 
The radial velocities of each source were derived using 
a two-dimensional cross-correlation function, using the zero-point 
established by the Earth's atmosphere as reference. 
More than 14 spectral lines in the CO $\Delta\nu$ = (0-2)
 bandhead window were used in the cross-correlation against 
LTE atmospheric models to compute the final results.
We found that the spectra of the protostars in our sample 
agree well with the predicted stellar photospheric profiles, indicating
that the radial velocities derived are indeed of stellar nature.
Three of the targets analyzed exhibit large radial velocity variations during the three observation epochs.
These objects - pending further confirmation and orbital characteristics - may become the first evidence for proto-spectroscopic binaries, 
and will provide important constraints on their formation.
Our preliminary binary fraction (BF) of $\sim$71\% 
(when merging our results with those of previous studies) 
is in line with the notion that multiplicity is very high 
at young ages and therefore a byproduct of star formation.}

\begin{document}
\maketitle
\section{Introduction}
\indent 
Since the early twenties that there is growing evidence that multiple stars 
may be the rule and not the exception.
This conception is more recentlt supported by observations of \cite{duq91} who found 
that the stellar multiplicity among main sequence (MS) stars in the 
solar neighborhood may be as high as $\sim$ 60\%.
Since then, more refined surveys probed the dependencies of the properties 
of stellar systems on physical parameters 
like age, mass, separation, and environment. These observations are highly relevant,
because there is still no 
comprehensive theory that explains the properties of multiple stars
 and their relation to star formation.
Recently, \cite{raga10} confirmed that stellar multiplicity 
decreases with stellar mass (or later spectral type) 
in the vicinity of the Sun, a finding that was already suggested earlier \cite[\textit{e.g.}][]{sieg05,lada06}. 
Because the stellar population in the solar neighborhood is dominated by late-type M-stars, this implies
that about half of the solar-like main sequence stars in the solar neighborhood occur in multiple systems. 
\\
\indent
It is likely that the multiplicity rate of stars is established in 
the early phases of star formation or
during the dynamical evolution that occurs afterwards 
\citep{good04, dedo04, bate09}.
And it is plausible that both initial and environmental conditions 
constrain multiple star formation,
and play a vital role in explaining their evolution over time, 
and their relation to the observed MS stellar multiplicity 
\citep{mat00,ster03,cone08,ko08,kacz11}. 
It is therefore essential to probe the stellar multiplicity as early as possible, 
 to gain a full understanding of the formation and the dynamical evolution of multiple systems. \\
\indent 
\cite{gh93,lein93,simon95,pat02,bec03} and \cite{duc04,duc07}
all noted differences in the BF of pre-main-sequence (PMS) stars in different SFRs. 
In particular, the multiplicity of T Tauri stars with ages $\lesssim$ 10$^6$ yrs, 
appears to be higher by a factor of 2 in 
less dense regions such as Taurus compared to denser SFR such as 
Ophiuchus \citep[\textit{e.g.}][]{simon95,bec03,ratz05} or the field \citep{pat02, duc04}. 
Toward even younger ages,
radio continuum \citep[\textit{e.g.}][]{rei04} and near- and mid- Infrared (IR) \citep{bars05, hai02, hai04, hai06} 
investigations found additional evidence for an even higher multiplicity among Class I and Flat-spectrum ($\sim$ 2 - 5 x 10$^5$ yrs) protostars.
\\
On the other hand, \cite{mau10} failed to detect companions in a small sample of 5 protostars at separations 50 \textless $a$ \textless 5000 AU ($a$ being the semimajor axis) in a millimeter study of self-embedded, young (Class 0) objects ($\sim$ 10$^4$ - 10$^5$ yrs). 
Taken together with the observations of \cite{loon00}, 
the authors argue that multiplicity in the separation ranges from 100 to 600 AU would only be defined in a later 
stage of star formation (namely after the Class 0 phase) and that early multiplicity in pristine systems may not be
as ubiquitous and primeval.
\\
\indent
However, all multiplicity studies in the earliest phases of stellar evolution are plagued with relatively small number statistics. 
In particular, very little is known about companions of embedded protostars at the sub-AU separation scales, a regime impossible 
to address with conventional imaging techniques. 
Spectroscopic Binaries (SB)  may provide additional important constraints on the star-formation mechanism itself.
\cite{toko02} and \cite{toko06} 
estimated the SB frequency in MS stars and inferred that 65\% of their sample of 165 spectroscopic binaries were
members of higher order multiple systems, often found in a hierarchical configuration. 
Remarkably, the frequency of SB in multiple systems strongly depends on the orbital period of the SB: 96 of the SBs with orbital periods
shorther than 3 days are in multiple systems, while this rate drops to 30\% for SBs with orbital periods longer
than 13 days. The higher frequency of spectroscopic system within triple or higher order systems
suggests an imprint of the formation mechanism itself. Spectroscopic pairs may have lost their orbital angular momentum
owing to the presence of a third body that interacts with the inner binary via Kozai cycles and
tidal friction \citep{kos72, ste05, fabr07}. This interaction has the effect of tightening the spectroscopic binary orbit, 
while the tertiary would eventually be evacuated to an outer region of the system. 
The SB may therefore hold an important fossil record of (prevailing) initial conditions during their formation.
However, this scenario is not conclusive by itself, and it is not clear at which evolutionary stage this process operates efficiently.
Observations are required to better constrain the detailed mechanisms of how tight binaries are formed at very young ages.
\\
\indent
The first successful PMS SB detections were performed by \cite{herbig77,hart86,mat89} and 
\cite{mat94} in the Taurus and Ophiuchus SFRs 
with a radial velocity precision limited to $\sim$1kms$^{-1}$ by optical high-resolution spectroscopy. 
Since then, improving instrumental capabilities and better radial velocity precision allowed to address PMS sources and the lower mass regimes  
\citep[\textit{e.g.}][]{mel03, cov06, ku06, joe06, joe08, prat07}.
Only recently, infrared (IR) high-resolution spectroscopy allowed the
 search for SB among the youngest, embedded PMS and the very low mass stars
\citep[\textit{e.g.}][]{prat07,bla07,bla10,fi10b,rice10,cro11}. In particular \citep[][]{fi10b,bla10,cro11} reached a precision of less 
than 50 ms$^{-1}$ in PMS stars and very low mass stars. 
Even the planetary mass regime may be in reach using high-stability, high-precision IR spectroscopy.
%\indent An additional motivation to study multiplicity in young and close stellar systems is to better understand its relation with planet formation \citep{egge07, egge10}.  Indeed, multiple stellar system strongly influence the circumstellar environment during the planet formation process  
% \citep{art94}. In particular the disk structure is affected by the presence of a stellar companion. 
%The migration of giant planets in the disk and their final orbital characteristics 
%depends on the distribution and dynamics of the masses of the entire stellar+circumstellar disk system.
%As planet formation is not a priori excluded in the presence of companion stars, the impact of companions on
%circumstellar disks require a more complete characterization of multiple stellar systems.
%
\\
\indent
Our own work aims to find SBs in a sample of 38 Class I/ II protostars in the SFR $\rho$ Ophiuchus.
We performed a high-resolution IR spectroscopic survey to derive precision radial velocities (RV) using the NIR spectrograph 
CRIRES at the VLT \citep{Kaeufl04}. 
We describe our sample, the data analysis methodology and results in Sect. 2. 
Sect. 3 discusses the RV derived in the context of the $\rho$ Ophiuchus SFR and its implications on the formation of SB at very young ages in general. 
In Sect. 4. we draw our conclusions.
\begin{figure}[ht]
\centering
\includegraphics[height=8.0cm]{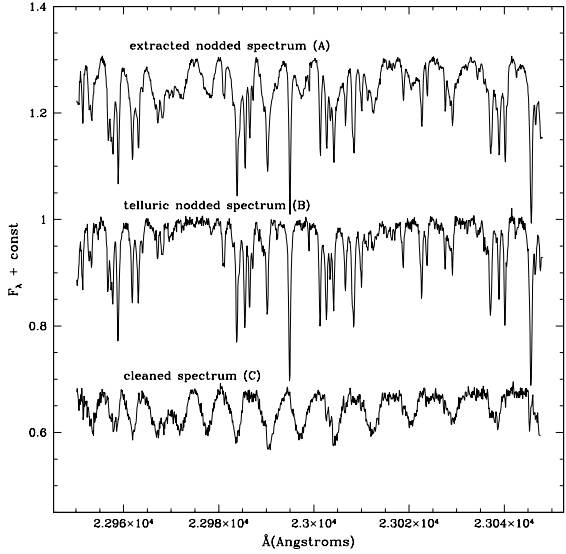}
\caption{Process of telluric signature removal in DET4. The extracted nodded spectra of GSS26 (A) is divided by the spectrum of the telluric standard stars with 
the same exposure time and observed closest in time (same airmass) (B). Lower panel exhibits 'cleaned' nodded spectrum (C).}
\label{allspec8}
\end{figure}
\section{Observational aspects and analysis}
\subsection{Observations and data reduction} 

The arrival of CRIRES \citep{Kaeufl04}, a high-resolution near- and mid-IR spectrograph 
mounted on  Antu at the Paranal Observatory allowed for the first time to acquire radial velocity
measurement precision of a few ms$^{-1}$ level in this wavelength window. CRIRES has opened the possibility to look
for (low-mass) spectroscopic companions around young stars that are otherwise too faint to be observed in the optical 
because of the extinction caused by circumstellar material. It also entails improved ability to detect dimmer companions
to optical bright primaries through more favorable flux ratios \citep[\textit{e.g.}][]{mace09}.

This study intends to take advantage of CRIRES to measure radial velocities in a sample of Class I/II protostars
whose multiplicity at larger scales has already been determined by imaging techniques. 
The main purpose consists in 
detecting radial velocity variations caused by potential close companions and to compare the overall 
multiplicity frequency in embedded sources as a function of separation with those 
observed for T-Tauri and MS stars from the sub-AU to a thousand AU scale. 
%
%%REFEREE POINT 1
%

\indent To this aim, a sample of 38 Class I/ Flat spectrum objects 
was selected from the near- and mid-Infrared multiplicity studies of \cite{hai02,hai04,hai06}.
The main selection criteria was the apparent $K$-band magnitude of the targets. As 
a compromise between the telescope time required to reach the desired S/R and the size of the sample, this limit was set at $K<11$. 
In addition, we selected proto-stars from 3 different SFRs to allow us a preliminary investigation of the role of the environment in our
measured spectroscopic binary fraction. 

%The factors taken into account for classifying their evolutionary status can be detailedly found there-in.
%Since light in protostars is mostly reprocessed in the circumstellar environment, 
%observations were conducted in the Near-IR most specifically in the K-band where  
%photospheric information could be accessed in the less embedded sources.\\
%\textbf{The criteria for selection were: 
%\begin{itemize}
%      \item magnitude in K band \textless11 to ensure detection using the dedicated CRIRES set-up (explained bellow)
 %     \item an homogeneous sample of both single and multiple sources;
 %     \item sources from different SFRs to enable multiplicity comparisons between different regions
%\end{itemize}
%}
%
%REFEREE POINT 2
%
In CRIRES the incoming light enters the instrument and is first directed 
into a pre-disperser, after which it is dispersed in a 31.6 lines/mm echelle grating. 
It delivers high-resolution spectra (up to $\sim$100 000 with the a 0.2''-slit) in the 960 - 5200 nm interval in wavelengths windows of $\sim\lambda$/70.
The observed light is finally projected into four Aladdin III detectors (4096 $\times$ 512 pixel) separated by $\sim$ 250 pix. 
The combination of the size of the detectors and the high resolution of the instrument limits the wavelength coverage of CRIRES, 
which is considerably smaller than that of  other spectrographs. 
Therefore the first task was to find out the best set-up among the 200 available to obtain the most accurate radial 
velocities as a function of the signal-to-noise ratio.\\
\indent  In order to estimate the S/R necessary to reach the required precision, we investigated the amount of information 
 (i.e., number of available spectral lines and their intensity) by simulating all CRIRES set-ups
  as described in \cite{fi10b}. In a nutshell, the estimated error on the radial velocity is given by
  $\epsilon(V_r) \propto R~[(S/R)<I>n^{1/2}_l]^{-1}$,
where R is the instrumental resolution, $\epsilon$ is the photon noise error, S/R is the signal-to-noise ratio, $<I>$ is the average depth
of absorption lines,  and $n_l$ is the number of lines entering in the computation of the CCF. In order to simulate all CRIRES set-ups, we
used the IR solar spectrum obtained with the Fourier Transform Spectrograph provided by the NOAO \citep{wallace93}.
This solar spectrum  was convolved with a Gaussian function and trimmed 
to match the resolution and wavelength
coverage of the various CRIRES set-ups \citep[for details see][]{fi10b}. 
Assuming
that only stellar lines deeper than a certain threshold contribute to the CCF, the above expression was applied
to all CRIRES simulated spectra and the quantity $\epsilon(Vr)/S/R$ was derived as a function of the set-up. 
This ratio allowed us to compare the radial velocity error (in a relative way) among the set-ups and to choose the most suitable one. 
A few set-ups were identified in the H- and K-band \citep[see Fig. 1 of][]{fi10b}. 
Because of the unknown degree of obscuration of our sample, we selected the K-band set-up.\\
\indent To 
derive the absolute error and therefore be able to transform the ratio $\epsilon(Vr)/S/R$
into exposure times, we had to know the absolute error on $V_r$ as a function of the S/R for one set-up. 
Our exposures times were then computed assuming that an S/R of 20-30 in the setting chosen here would give a 150m/s error for the set-up { former \tt 24/-1/i setting}.
In practice, this estimate turned out to be too optimistic. As a consequence, exposure times for several targets were underestimated.
%
%\indent Unfortunately, we were unable to detect the stellar signal
 %in most of the stars proposed for observations. \textbf{The reason for this result was because the proposed exposure
 %times were insufficient for integrating protostellar light. Despite the clarity of ESO ETC tool it was difficult 
%to predict the final depth of the stellar lines using our CRIRES configuration. It was the first time a resolution blabla
%To run the experiment completely, new observations will be proposed in the near future.}  
%
Accordingly, we obtained 15 protostars with robust RV measurements from the original 38 stars proposed and only 
7 sources with multiple RV measurements at sufficient S/R. This paper focus on 
these 7 multi-epoch targets from the $\rho$ Ophiuchus SFR, for which we detected photospheric absorption features. 
In Table~\ref{table:1} we present the properties of the full sample.
An assessment of the evolutionary status of each sources is presented in the column SED of Table~\ref{table:1}.
It follows the usual YSO classification based on the  
change of the slope of the SED as a function of age. Rising SEDs with positive slopes toward mid-IR usually depict the presence of
 protostellar envelopes that are generally more massive than the 
protostar. These protostellar envelopes tend to disappear in the subsequent phases of evolution, namely the Class I phase.
As a consequence, the SED slope flattens (flat-spectrum sources) and then becomes negative towards the Class II or T-Tauri phase.
\citep[\textit{e.g.}][]{moan01}.\\
%were taken from the same references and are based on the spectral indexes derived for these sources in the mid-IR \citep[see \textit{e.g.}][]{hai06}. 
%\textbf{Just for clarity,} SED classification is an empirical measure usually used to assign an evolutionary stage to YSO.
%
%
%
%(Pedro Tabela 1, we should have the collumns: RA,DEC, SIMBAD identifier, Date Obs, DIT, K, Nodding cycles, SN, SED, Rer for SED, SFR)}
%
%More information on each target can be found in Table~\ref{table:1} \textbf{and in different studies produced over the years 
% 
%
%\indent The results presented in this paper were derived using high resolution spectra of \textbf{these} 7 protostars with 
%detected photospheric absorption features located in the $\rho$ Ophiuchus star forming region 
%and observed by \cite{hai02,hai04,hai06}. 
\indent Our sample was observed between April 2008 and February 2009 with a set-up
 covering a wavelength range of 2254.2 to 2304.7 nm. 
The center wavelength reference in detector3 (DET3) was $\lambda$2287.1 nm.
The CO (2-0) bandhead at 2293.5nm was registered on detector 4 (DET4).
Owing to the lack of reference stars close to our science targets, our observations were carried out in the seeing-limited mode. 
Observations were collected under good seeing ($\sim$ 0.8") in 2 nodding  cycles (i.e., 4 total independent exposures) \footnote{In one nodding 
cycle two exposures are taken in total, one at position A and another at position B. In the second nod cycle, a third exposure is taken again at position B, 
then the telescope is moved back to position A where the 4th exposure is taken.},
with a 0.3'' wide slit allowing a resolution of $\sim$ 60,000 which we found to be the best compromise between the amount of flux entering the
spectrograph and the ability to resolve the lines of the object and those from the atmosphere.
Final S/N ratios given in Table~\ref{table:1} were computed on small featureless regions.\\
\indent The approximate range of masses of our 7 protostellar targets was taken from \cite{natta06} and  
assumed to be within the interval from 0.18 $M_{\odot}$ to 1.4 $M_{\odot}$.  Assuming that semi-amplitudes of 500m/s
can be measured at a 3$\sigma$ level, we are able to detect companions ranging from several jupiter masses with periods of one or a few weeks 
to equal-mass companions (from brown dwarfs to solar-type stars) with periods of several years.
A more detailed study of the detectability rate similar to the one performed by \cite{mel03} 
will be carried out at the end of the survey.\\
%
%\indent A short comment should be added to the sources we were unable to detect in our near-IR observations. 
% Several reasons can be at the origin of this outcome.
%One of them could be that at the time of the first visits to each target, in April and May of 2008, a loose tape in CRIRES was blocking the entrance of the slit 
%therefore creating a strong vignetting of the incoming light. Another source of explanation is 
% that extinction and the intrinsic YSO nature of our sources demanded a higher exposure time to integrate stellar light.
%If, for instance, we face sources with considerable variable extinction, 
%we may have to assume that the exposure times, computed with ESO's online \textit{exposure time calculator} tool 
%on the basis of the stellar magnitudes and spectral indexes as indicated in \cite{hai04, hai06}, could have been underestimated. \\
% 
\indent In addition to the protostellar data we also collected spectra of 
telluric standard stars (featureless early-type B stars)
observed with the same airmass and instrumental setup as our science targets. 
Multi-epoch spectra of 3 radial velocity standard stars 
were collected during the timespan of our observations 
in order to study the precision of our RV measurements. \\
%Finally, observations of spectral type standards were also performed in the course of our program to better constraint 
%the spectral type of our targets. \\
%
%\input{tabla_vrad_nodded}
%
\indent Data were reduced using an optimized IRAF-based pipeline to ensure a uniform and homogeneous reduction for all targets in the different epochs. 
The reduction steps in our pipeline were: 1) correction for CRIRES nonlinearity 
using ESO on-line coefficients and construction of bad-pixel masks, 
2) dark-current subtraction from the flatfield images and collapsing into a master flat, 3) division of science spectra 
by these final flatfield images, 4) subtraction of opposing nodded spectra for 
dark-current removal, 5) optimal extraction using the \cite{h086} algorithm, 6)
telluric signature removal by dividing each extracted spectrum by the spectrum of the telluric STD stars with the same exposure time and 
observed closest in time (same airmass) (Fig.~\ref{allspec8}). \\
\indent 
The position of the CCF peak strongly depends on the average center 
of all stellar lines used to computed the CCF.
Because telluric removal could have changed the final line shapes, it is essential to find the most similar
and adequate telluric spectra so as not to change the protostellar line shapes fundamentally. 
A test was carried out to assess the impact of the use 
of different observed telluric spectra in the derived RVs. 
First, we performed for a specific date telluric line removal
 with the observed telluric standard and the combined science spectra 
of protostars where no stellar signal was detected (i.e.,
only the atmospheric lines are imprinted in the data). 
The main conclusion was that small
fluctuations ($\lesssim$ 50ms$^{-1}$) on the final RV could indeed be detected. 
These discrepancies were found in tellurics
that were observed $\geq$2 hours after the science target. 
We found no relevant differences in the final RV when the telluric removal was performed with tellurics observed within that time difference. 
We conclude that telluric removal as explained here has no major impact on the final RV. \\
\begin{figure}[t]
\centering
\includegraphics[scale=0.45]{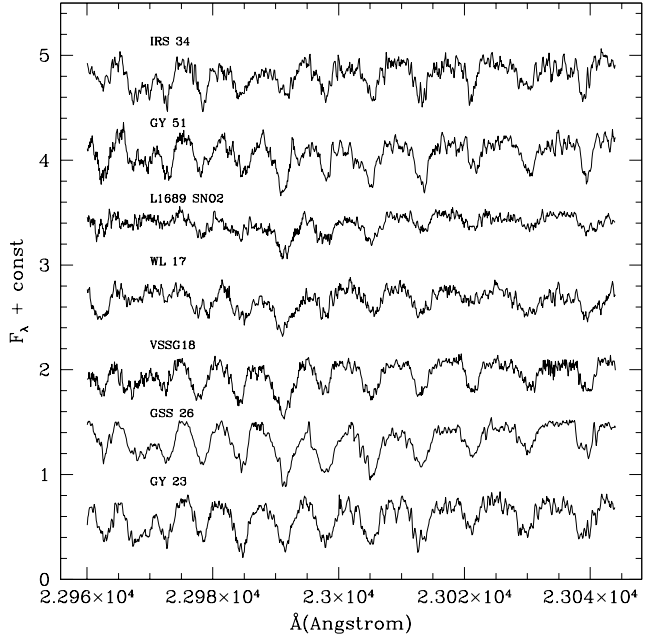}
\caption{Extracted spectra of our Class-I stars as observed in Detector 4. The data shown here were
 smoothed for clarity, continuum-normalized and Doppler-corrected to a null radial velocity.}
\label{allspec}
\end{figure}
\indent The extracted spectra were wavelength-calibrated using telluric absorption lines.
 The use of telluric features to build a wavelength solution 
is now a well-documented technique \citep{cac85, bla07, fi10b, bla10, cro11} and it is highly convenient  
when gas cells are not available as was the case of CRIRES at the time of the observations. 
We refer the reader to \cite[\textit{e.g.}][]{fi10a} or \cite{bla10}
for an extensive explanation of the robustness of this technique. 
The study of \cite[][]{fi10a}, in particular, 
showed that O2 atmospheric lines can be stable down to 10 m/s over long periods of time (6 years). \\
\indent Basically, wavelength solution construction consists in finding and fitting (in each observed spectrum)
the center of telluric lines with low-order polynomials to match 
their theoretical value as given by online databases. 
In our particular setting, we took advantage of the wealth of CH$_4$ atmospheric 
telluric lines to build suitable wavelength solutions for each 
extracted spectrum. This set of CH$_4$ lines has already been used 
in previous studies and was shown to deliver precise RVs 
in very low-mass stars and in brown dwarves \citep{bla07,bla10}.
The reference wavelength of each spectral line was taken from 
the HITRAN database with a typical accuracy $\lesssim$ 50ms$^{-1}$,
 which is well-suited to our purposes, giving the radial velocity 
amplitudes expected by the presence of stellar companions.
  Spectral features of protostars are known to be considerably broader
 than MS stars or even class-II stars \citep{gre97, dop05},
 because of their typically high \textit{vsini}. For this reason, 
we expected to find a high level of blending between spectral features 
of telluric and stellar origin that could introduce 
significant systematic errors when building the wavelength solution or determining the RV measurements. 
To minimize possible systematic errors arising from the overlap of the telluric with the stellar signal, 
we combined all nodded images from each target's observation
into a single frame for building the wavelength solutions. This procedure considerably 
improved the S/N of the final spectra, which is therefore
ensured a better identification and fitting of telluric lines and their theoretical centers. 
This steps assumes that the
 spectral characteristics of the atmosphere did not vary between nodded positions (i.e., in time-scales of $\le$4 min in K-band).\\
\subsection{Data analysis}
\indent The RVs in this study were derived using a slightly different version of the \cite{fi10b} pipeline, 
a two-dimensional (2D) cross-correlation function (CCF)
 based on TODCOR  \citep{maz92}. The pipeline of \cite{fi10b} was specifically designed to deliver the RV of the target 
relative to the zero-point established by the Earth's atmosphere. \\
\indent In a 2D cross-correlation using atmospheric lines as a reference, the RV of the atmosphere (RV$_{sky}$) and the RV of the star
 (RV$_{star}$) were obtained from the same spectrum by cross-correlating the observed spectrum with a combination of two CORAVEL-type masks 
\citep[i.e., the absorption lines are represented by box-shaped lines, see][]{bar96}, one 
 mask containing the telluric lines and another the stellar lines.
 As mentioned in the previous section, the lines present in the atmospheric mask use 
the HITRAN wavelengths as reference. The stellar masks were built
 using the synthetic spectra of PHOENIX for stellar type G8V, K2V, K5V, M1 and M6. The width of the box-shaped line was set by
 convolving the synthetic spectra with Gray's rotational profiles \citep{gra92} with the rotational velocities chosen to adjust the line width of the observed spectra. 
The limb darkening coefficient was found to be irrelevant so we assumed a standard value of 0.6.

We stress that different combinations of spectral type (i.e., $\log g$ and $T_{eff}$), 
rotational velocity and veiling would well fit the observed spectra equally.
 For this reason, the process above was useful to set the line center and width 
and to generate the box-shaped mask that was used in CCF computation. 
No attempt was made to derive the target's $T_{eff}$ and $v\sin i$. 
 To test the sensitivity of our radial velocities with the stellar mask, we built a new series of stellar masks 
using the signal-to-noise spectrum of our RV standards HD129642 (K2V), Gl406 (K5V) and HD 105671 (M6). 
We found that changing the spectrum to build the masks induced RV variations below 
100 m s$^{-1}$, which is within our photon noise. 
 %\textbf{Each mask is composed by a series of holes where the spectral lines of the selected origin 
%are supposed to be found in the case the object of interest has RV equal to 0. Since, in our case, both the atmosphere and the source 
%have a RV different from zero, we should calculate the 2D CCF.} 
%This 2D function gradually takes shape when both masks are shifted in RV within a chosen velocity window.

\begin{figure}[t]
\centering
\includegraphics[scale=0.7]{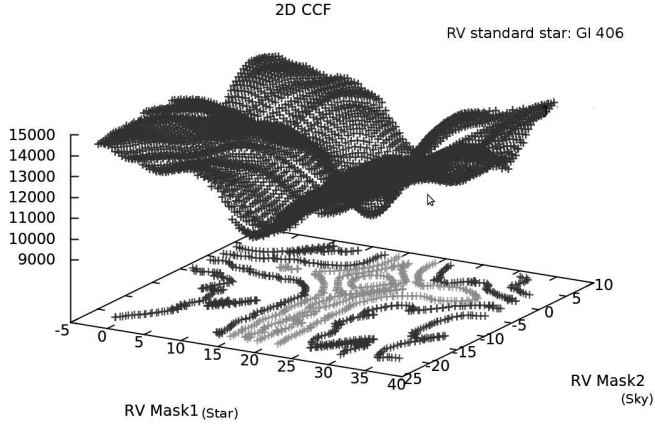}
\caption{Resulting 2D CCF for the star Gl406 and telluric lines 
 using an M6 stellar mask and an atmospheric mask, respectively.
The best correlation is found for the atmosphere (RV Mask2) and the star (RV Mask1) 
at the lowest point of the function in the z-axis. }
%As we can see in this particular case, 
%the RV$_{star}$ was found to be $\sim$24.06kms$^{-1}$ and the RV$_{sky}$ $\sim$6.61kms$^{-1}$}.

\label{2dccf}
\end{figure}

The best solutions for the RV$_{sky}$ and RV$_{star}$ were found
by fitting the 2D CCF surface (as shown in Figure\ref{2dccf}) with a two-dimensional Gaussian function. 
The coordinates of the minimum give RV$_{sky}$ and RV$_{star}$.
\citep[for a description of the 1D CCF technique see][]{bar96}.
  
In our specific case, the most important aspect of our adapted pipeline compared to the original 2D CCF of 
\cite{fi10b} is that it uses two spectra: the one produced after the telluric removal (\textit{e.g.} spectrum C 
in Fig.~\ref{allspec8}) 
when computing the stellar CCF against a stellar mask, and the original one containing the stellar and the telluric lines (\textit{e.g.} spectrum A in the same figure) 
when computing the atmospheric CCF against an atmospheric mask. Although it uses two spectra, 
the process is still performed simultaneously, so that the final product is still a 2D function of the velocity shifts of the two templates (see Fig.~\ref{2dccf}).
%The CCF of a given set of RV$_{star}$ and RV$_{sky}$ is, therefore, the sum of the CCF of the original spectrum (correlated against a telluric mask) with 
%the CCF of the cleaned spectrum (correlated against a stellar mask).
%This technique ensures the acquisition of the RV$_{star}$ relative to the telluric atmospheric scenario at the time of the observations.  \\
All wavelength solutions and derived RVs were corrected for the solar system center of mass following the references given by \cite{bret88}. 
Barycentric Julian dates were also retrieved according to the same source. 
The final stellar velocity is then given by $RV_{final,star}=RV_{star}-RV_{sky}+BERV$, BERV being the barycentric Earth's radial velocity 
on the UT date on which each target was observed.

%
%\indent For targets whose individual nodded spectra had enough S/R
%RVs for each target were determined by cross-correlating each nodded spectrum against 
% spectra provided by the PHOENIX stellar models \citep{hausc99,bar05} and against the
% spectra of the RV template M6V star Gl406 observed with CRIRES. 
% All wavelength solutions and derived RVs were corrected for the solar system center of mass following the references given by \cite{bret88}. 
%Barycentric Julian dates were also retrieved according to the same source. \\ 
%
\indent We note that during telluric removal the final S/N of the cleaned spectrum (C in Fig.~\ref{allspec8}) decreased 
to about $\sim$20. The S/N of individual nodded positions was often low. Therefore, 
with the exceptions of the sources GSS26 and GY23 and the RV standards, we combined the nodded spectra
to form a single spectrum in order to increase the final S/N. 

%In any case, and taking into account the noise affecting the background continuum level of the spectra,
In addition, we also included an algorithm (analog to the continuum task from the IRAF package) 
to better fit the noisy continuum around each spectral line used in the cross-correlation. 
This last procedure had the effect of ''cleaning'' the final shape of the CCF from spurious 
spatial frequencies introduced by the incorrect fitting of the continuum level in the noisiest spectra. 
%
%\indent While the chief aim of this work was to compute the RV of the observed protostars, it was also relevant, however, 
%to find the tools used to deliver results with the highest precision possible. In this context, all factors affecting precision
%such as the best models used to establish cross-correlation, the process of telluric removal, the \textit{vsini} of the protostars had to be thoroughly studied,
%their effects upon the RV measurements quantified. \\
%
\subsection{Results}
\begin{figure}[t]
\centering
\includegraphics[scale=0.45]{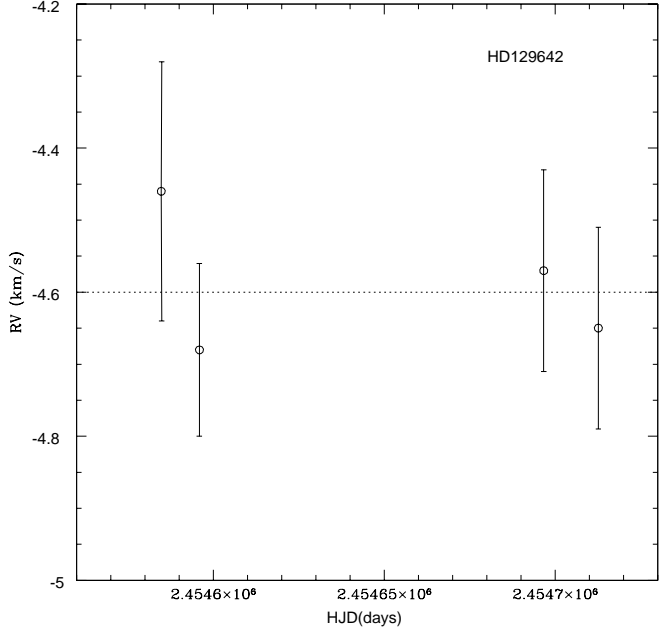}
\caption{RV measurements for the STD star HD129642. Error bars come from the relation \textless$\sigma_{vrad}$\textgreater * N$^{-1/2}$
 being $\sigma_{vrad}$ the dispersion of the RV determined in each nodded image and N the number of individual measurements.}
\label{allspec3}
\end{figure}

\indent The RV measurements of stars that possess multi-epoch observations are shown in Table~\ref{kstars} 
and are plotted in Fig.~\ref{allspec3} and \ref{allspec4}. Final RVs come 
from the 2D CCFs calculated using the spectral information
in the wavelength region of the CO $\Delta\nu$ = (0-2) bandhead window 
 sampled in DET4 in the chosen CRIRES setting ($\sim$ 14 absorption lines). 
%This bandhead region has an extended and fairly known periodic structure and as a consequence 
%it is a good indicator of rotation \citep{gre97} since its shape is strongly dependent on the \textit{vsini} of the parent source.\\
%
%\input{tabela_Teff_vsini}
%
\indent 
The errors quoted in Table~\ref{kstars} were computed as follows. 
For each independent exposure (nodded spectrum), an RV 
was derived for each of the 14 lines in the region of the CO $\Delta\nu$ = (0-2) bandhead. 
A weighted mean was calculated using the line depth as weight. 
%The deeper the feature the more spectral information it has, 
%so the higher is its individual contribution to the final CCF. 
An average deviation of the 14 RVs was then computed.
The final $\sigma$ quoted in Table~\ref{kstars} is then the average $\sigma$ 
divided by $N^{1/2}$, where $N$ is the number of the total nodded spectra. 
Errors for GSS26, GY23 and the RV standards (HD129642, HD105671, Gl406) 
were computed as described above.
For the remaining targets where we had to combine all available exposures, the errors presented in Table~\ref{kstars} 
are the RMS of the RVs computed using the 14 lines.
The errors range from around 0.015 (in GSS26) to 1.20 kms$^{-1}$ in the noisier stars (S/N$\sim$15).\\ 
We caution that the low errors obtained for some stars in Table~\ref{kstars} are probably underestimated. 
According to \cite{fi10a} and \cite{fi10b}, the precision one can 
obtain using the telluric lines as zero-point is better than 10m s$^{-1}$. This is about 
what we obtain on the RV standards. On the one hand these 
low uncertainties are telling us that the systematic errors 
(target centering, detector characteristics, etc.)
are not dramatically higher than those fund on the time-scale of our measurements (100 days or so). 
On the other, they might simply reflect the low-number of measurements obtained for our targets. 
A more realistic estimation of the errors will therefore be obtained later in our survey.

%
%\indent   by estimating the relation of the standard deviation with the \textit{vsini} and the 
%S/N of the spectra \citep{bla07}.  
%In more detail we added different gaussian noise seeds (to reach the S/N $\sim$  20-30 of our combined spectra) 
%to one of the rotationally broadened synthetic model (with parameters T$_{eff}$~=~3649K, \textit{vsini}~=~20 kms$^{-1}$) and produced theoretical spectra with 
% S/N that ranged from 2 to 100. We then doppler shifted these spectra with values of RV in a linear interval of velocities from
% -20kms$^{-1}$ to -5kms$^{-1}$ to match the range of velocities found in the sample spectra. 
%Final dispersion was determined by deriving the parameters of the noisy models with the same S/N, \textit{vsini} and RV of our observed spectra. 

%
\indent To assess the precision obtained by the method presented here, we computed the RV for the RV standards Gl406, HD129642, and HD105671. 
%and compared our results to precise results ($\lesssim$0.120 kms$^{-1}$) provided  by the literature. 
For Gl406, we obtain 19.730 kms$^{-1}$ with an internal 
dispersion of 0.010 kms$^{-1}$ in all nodded images. 
Our results is very close to the published result provided in \cite{wb03} of 19.175 kms$^{-1}$
and to the 19.18$\pm$0.03 km s$^{-1}$ obtained over 11 measurements (Forveille, private communication). 
Quite striking is the RV = -4.59 kms$^{-1}$ and the low dispersion ($\sim$0.08 kms$^{-1}$) obtained in the four different epochs in the star HD129642. 
Results are plotted in Fig~\ref{allspec3}.
An RV precise value of -4.60 kms$^{-1}$ is given by \cite{gon06}. The four epochs obtained for HD129642 coincide  with
the temporal window of most of our Class I/II sources. That they are able to reproduce HD129642 RV during such an extended period 
confirms the stability and precision of the results. \\
%
%
%\indent It is hard to quantify the final level of precision delivered by our RV measurements. Systematic errors arise from different quadrants of our analysis,
% for instance in the: 1) wavelength solution construction, 2) determination of the RV of the telluric lines, 3) telluric removal, 4) definition of the continuum levels 
%of each spectra line, 5) choice of the models to cross-correlate and their physical characteristics. The work with the STD revealed that the precision can be variable 
%depending on the conditions of the spectra, namely the S/N. Such dependence make us understand that the simulations with the synthetic spectra, described above, 
%are the best approach to infer the standard deviation and constraint the relevance of the absolute RV values found. 
%
\section{Discussion}
\subsection{Origin of the CO: photospheric or circumstellar}
\indent Protostellar light is often reprocessed in the circumstellar 
environment and becomes contaminated with emission or absorption from different origins.
Particularly in the near-IR the circumstellar disk is thought to play a central role 
\citep{cal91} in altering the stellar signature
because the temperatures of the excited material emit in the same wavelength as the protostellar photosphere. \\
\indent For the CO regions, for instance, dedicated studies in TTS with accretion disks 
\citep[\textit{e.g.}][]{carr93, naj96, lu99} have uncovered observations of CO overtone bandheads.
The presence of inclined circumstellar disks can affect the shape of observed CO spectral lines that can exhibit anomalous 
double peaking profiles or depict a wider velocity field gradient.
%(broadening of the CO lines can be $\geq$100 ms$^{-1}$).
These features can be observed mainly in emission, but can 
also be found in absorption \citep[see \textit{e.g.}][and references therein]{hoff06} 
and are apparently associated with the most active PMS sources 
such as the Fu Orionis objects \citep{hart04}. 
CO absorption lines can be produced in a circumstellar disk when its temperature 
profile depicts a negative vertical gradient very much like a stellar 
photosphere \citep[see \textit{e.g.}][]{kauf05}.  \\
\indent Nevertheless, because the dynamics of the material 
in circumstellar disks present Keplerian velocity profiles \citep{casa96},
 the CO absorption of circumstellar nature drastically contrasts with the 
CO absorption profiles of photospheric origin. 
High-resolution observations are usually able to separate both components.\\
\begin{figure}[t!]
\centering
\includegraphics[scale=0.45]{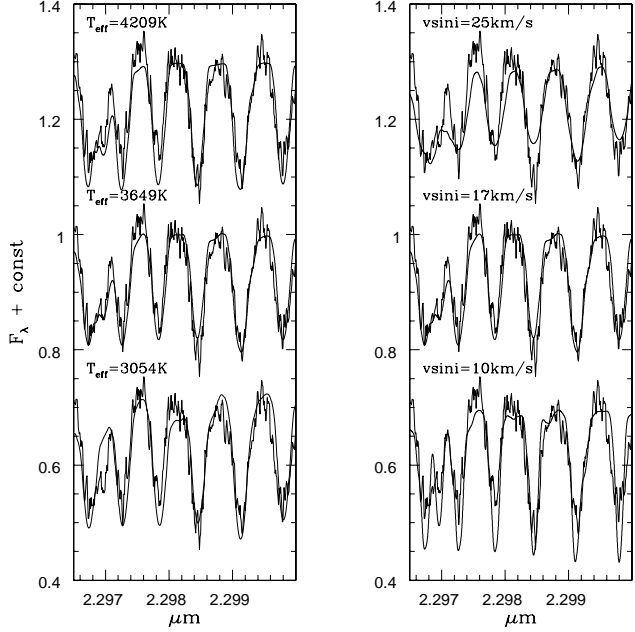}
\caption{Observed protostellar spectra that we were able to reproduce with rotationally broadened stellar photospheric profiles.
Right panel: Overplotted are some synthetic spectra (with a fixed T$_{eff}$ of 3649K) (solid lines) with different \textit{vsini} rotations
with the spectrum of the protostar GY23 on one of the dates it was observed. Left panel: 
The same, but with a fixed \textit{vsini} of 17kms$^{-1}$ and different 
temperatures.}
\label{allspec2}
\end{figure}
\indent In Fig. \ref{allspec2} we show that our protostellar spectra can be reproduced by 
rotationally broadened photospheric synthetic spectra. As an example in this figure, 
we rotationally broadened 5 solar metallicity PHOENIX stellar models of Spectral 
type G8V, K2V, K5V, M1 and M6 (T$_{eff}$ ranging from 5650K to 3050K) 
 with the same resolution as our observations and overplotted them with the spectrum of the source GY23. 
The match between the synthetic and the observed profiles is evident. 
Using different sets of models we were able to extend
 this result to all 7 protostars presented in this work.
For this reason we believe that the lines analyzed here are generated in a 
photospheric environment and that the RVs found are of stellar origin.
%Moreover, the small values of the CO line equivalent widths ($\lesssim$ 50 kms $^{-1}$) of our observations provide a clear indication 
%that these lines are of photospheric origin and not keplerian.  \\ 
%
\subsection{Radial velocities}
\addtocounter{table}{0} 
\setlength\extrarowheight{2pt}\vspace*{0.5pt}
\begin{table*}[htbp]
\caption{Barycentric Julian dates of each target, derived RVs, and final dispersion for each date it was observed. 
In the upper table the results from each nodded image are presented. In the lower 
panel we show the RV measurements in images of sources that yielded an insufficiently 
S/N after telluric removal and that had to be combined for better precision.}
    % title of Table
\label{kstars}      % is used to refer this table in the text
\centering                     % used for centering table
\begin{tabular}{cccccccc}        % centered columns (4 columns)
\hline\hline                 % inserts double horizontal lines
Source & MJD &v\textit{rad} [km s$^{-1}$]  & \textless$\sigma$\textgreater & Source & MJD & v\textit{rad} [km s$^{-1}$]& \textless$\sigma$\textgreater \\
\textit{nodded images}\\
\hline
%\hhline{-~~~~~~~~~~~~}
%
GSS 26  & 2454584.63565 & -6.97198&  & HD129642 &  2454584.62788 & -4.48490&\\
& 2454584.63909 & -6.93852&  &  & 2454584.63176  & -4.59242 &\\
& 2454584.63745 & -6.94756&  & &2454584.62387 & -4.22596   & \\
& 2454584.64087 & -6.96722 & 0.01 &  & 2454584.61976  & -4.59242& \\
& 2454595.81134 & -6.93863 &  &  & 2454584.63573  & -4.60915 &\\
& 2454595.81313 & -7.03562 &  &  & 2454584.63931  & -4.26827 &0.09\\
& 2454595.81492 & -6.95636 &  &  & 2454595.76527  &  -4.48264 &\\
& 2454595.81664 & -6.89839 & 0.03 &  &  2454595.76193 &  -4.71534&\\
& 2454700.50155 & -6.98628 &  &  &  2454595.75729 &  -4.67921 &\\
& 2454700.50334 & -6.99655 &  &  &  2454595.75303 &  -4.64180&\\
& 2454700.50501 & -6.94632 &  &  &  2454595.76937 &  -4.72930&\\
& 2454700.50681 & -6.86636 & 0.03 &  &  2454595.77303 &  -4.82873&0.06\\
GY23 & 2454584.66269 & -11.74129 &  &  & 2454696.51517  & -4.59595 &\\
& 2454584.66334 & -11.91061 &  &  & 2454696.51934      &  -4.49518&\\
& 2454584.66383 & -11.61631 &  &  & 2454696.51108  &      -4.63937 &\\
& 2454584.66435 & -11.92826 &  0.07 & & 2454696.50713  &      -4.34328& \\
& 2454595.80848 & -6.86636 &  &  & 2454696.52319  &      -4.60013& \\
& 2454595.80786 & -6.99655 &  &  & 2454696.52733  &      -4.75527& 0.07\\
& 2454595.80737 & -6.94632 &  &  & 2454712.50231  &     -4.80776 &\\
& 2454595.80672 & -6.82011 & 0.04 &  & 2454712.50480  &     -4.59263 &\\
& 2454710.51703 & -10.14933 &  &  & 2454712.50349  &     -4.68273&\\
& 2454710.51638 & -9.97590 &  &  & 2454712.50343  &     -4.72138 &\\
& 2454710.51590 & -10.23329 &  &  & 2454712.50391  &     -4.44132&\\
& 2454710.51526 & -10.11078 & 0.05 &  & 2454712.50429  &     -4.62820& 0.07\\
Gl 406  & 2454890.64318 & 19.73374& &  HD105671 & 2454795.13436 & -4.20193 &\\
& 2454890.64215 & 19.72979 &  &  &  2454795.13447 &   -4.22028  & 0.01\\
& 2454890.64125 & 19.74171 &  &  &  &   & \\
& 2454890.64024 & 19.70016 & 0.01 &  &   &  &  \\
\hline
\hline
\textit{combined images}\\
\hline
GY 51  & 2454584.67379 & -10.75198 & 0.22 & L 1689 SNO2  & 2454699.66023  & -7.76361 & 0.76\\
& 2454595.83890 & -9.44251  & 0.35 &  & 2454873.87196 &  -7.26417  & 0.56\\
& 2454597.72588 & -6.26190 & 0.25 &  & 2454884.87905   &  -5.76223  & 0.92 \\
& 2454710.527356 & -10.18371 & 0.73 &  &   &      \\
VSSG 18   & 2454597.88737 & -15.89123 & 0.28 & IRS 34  &  2454597.80272 & -7.36514 & 0.63\\
& 2454612.72451 & -13.73072  & 0.32 &  & 2454612.80033 &  -7.17129  & 0.80\\
& 2454733.55090 & -17.12639 & 0.44 &  &  2454712.56017 &  -7.25198   & 1.22\\
WL 17  & 2454597.75984 & -5.62187 & 0.65 &   &   &  &\\
 & 2454610.76480 & -6.12428 & 1.10 &   &   &  &\\
\hline %inserts single line
\end{tabular}
\end{table*}

\indent From the RV measurements presented in Table~\ref{kstars} we can see that the range of values obtained (from $\sim$ -5.5 to -17 kms$^{-1}$) 
 seems to indicate that the sources studied here are indeed from the $\rho$ Ophiuchus cloud. 
All targets, with the exception of VSSG18, are within 3~$\sigma$ of the characteristic velocities exhibited by $\rho$Oph sources, -6.3 $\pm$ 1.0 kms$^{-1}$ 
\citep{ku06, prat07}. \\
\indent The phenomenon of accretion and stellar activity in protostellar ages has the effect of 
altering spectral line-shapes.
These non-periodical variations change the line-shape and
induce RV variations that can reach amplitudes of the order 
on the $\sim$2-3 kms$^{-1}$ \citep[\textit{e.g.}][]{ale05}.
%In this context, the RV variations of that order of magnitude could be a by-product of accretion instability.  
%Fortunately, the effects of this accretion-driven RV modulation do not persist for long (see \textit{ibid}). 
%Most probably this phenomenon in conjunction with the uncertainties driven by the 
%low signal-to-noise ratio of the data from IRS34, L1689SNO2 and WL17 are the responsible for the small variations found. \\
%
Therefore, although it is possible that some of the observed RV variations might be caused by
 stellar activity and accretion, the amplitude of the variations ($\simeq$ 4-6 kms$^{-1}$) observed for  VSSG18, GY23, and GY51
suggests that those variations could have been caused by a spectroscopic companion.
%It is possible that the overall RV differences uncovered display our inability to accurately address the physical properties of Class I/ II sources
%(\textit{e.g.} stellar activity and accretion). 
%
\begin{figure}
\centering
\includegraphics[scale=0.45]{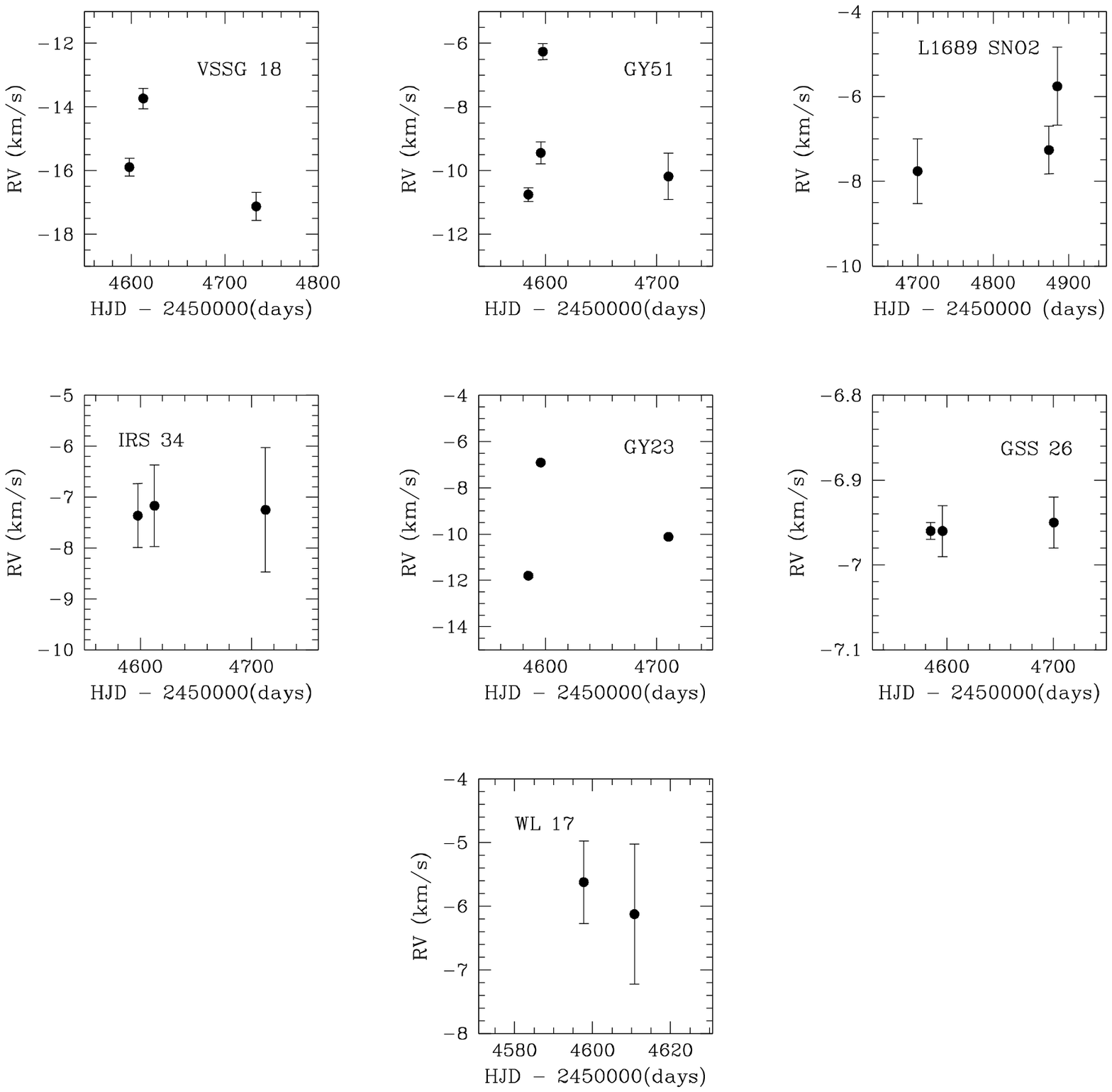}
\caption{Radial velocity measurements of the 7 protostars of our sample. Error bars in GSS26 and GY23
 are calculated from the relation \textless$\sigma_{vrad}$\textgreater * N$^{-1/2}$
 being $\sigma_{vrad}$ the dispersion of the RV determined in 
each nodded image and N the number of individual measurements. In all others targets 
 error bars were computed as explained in Sect. 2.4..}
\label{allspec4}
\end{figure}

Moreover, for the other 3 proto-stars  observed with lower S/R (IRS34, L1689SNO2 and WL17) low-amplitude RV variations were detected,
suggesting that the RV variations found among VSSG18, GY23, and GY51 might well be produced by companions.
Because the effects of this accretion-driven RV modulation do not persist for long (see \textit{ibid}), the only
way to confirm the nature of the observed variations is to assess the eventual variations of the orbital parameters. In particular, variations
of the semi-amplitude and/or the phase would indicate  that the RV modulations are likely to be accretion-driven 
\citep[e.g.][]{hue08}.

\indent Finally, concerning VSSG18 alone, all RV measurements are well above the 3 $\sigma$ of the mean of the velocity distribution of $\rho$Oph from \cite{ku06}. 
VSSG18 average velocity is of -15.873 kms$^{-1}$, a clear but not extreme outlier from the local cloud. 
Based on its spectral energy distribution and spectral index from previous imaging studies
it seems that VSSG18 is indeed a proto-star. Its deviant RV could suggest that this is long-period SB1 system.
\\
\subsection{Multiplicity and star formation scenarios}
\indent The use of CRIRES along with the technique described in the present work allowed us for the first time to carry
out a systematic search for spectroscopic companions at early stages of their formation (below about 1Myr or so).
Although preliminary and covering a small number of objects 
(7 objects only), the present paper  brings already some interesting aspects concerning the 
configuration where those systems are found.

Out of 7 sources, 4 are known to be in multiple visual systems, namely, L1689NO2 \citep{hai04}, IRS34 
\citep{bars05}, GY51 \citep{hai06}, and GY23 \citep{eli78,hai02, hai04},
whereas the other 3 targets (GSS26, WL17 and VSSG18) 
did not have any detected companions in previous imaging surveys \citep{hai04, hai06, bars05, duc04}.

GY51 and GY23 resemble some young multiple systems found in the literature. The former, is maybe a non-hierarchical system such as HD~34700 \citep{ste05} 
whereas the latter looks like to be in the typical hierarchical configuration with a tight inner binary with an additional companion 
which probably caused the inner orbital evolution of the inner pair \citep{toko06}. 
 
The detection of a potential spectroscopic VSSG18 system without visual companion detected is interesting. 
If we believe that one possible (or maybe the only possible) 
way to form spectroscopic pairs is through the existence of an additional companion, 
systems such as VSSG18 are valuable because, if confirmed, they might set time constraints for orbital 
evolution to occur. Also, they could be interesting targets 
for AO-assisted campaigns to look for more close companions.

\indent When we combinethe results of our study with those of 
\cite{hai04, hai06}, \cite{bars05}, and \cite{duc04},
there were 2 single systems (S) amongst the YSO analyzed systems, 3 binaries (B), 1 triple (T) (1~spectroscopic binary~+~1visual companion), 
and one quadruple (Q) (1~spectroscopic binary~+~3 visual companions). The binary fraction (BF) of this small sample of protostars ($B+T+Q)/~(S+B+T+Q$) is 
 $\simeq$71\%. This preliminary result, although based on such a small sample, 
agrees surprisingly well with the notion that multiplicity is very high at young ages and therefore
it might be a product of star formation. We advise caution in the interpretation of this final BF, however, because it 
is likely affected by low number statistics and it requires confirmation with a larger sample of protostars. \\
\section{Conclusions}
\indent We have performed a spectroscopic multi-epoch survey of 7 known
protostars in the $\rho$Ophiuchus star-forming region in the near-IR. 
Our main goal was to derive the close-binary frequency of embedded sources
where multiplicity is already known from imaging techniques.
We successfully derived radial velocities of Class I/II sources with an unprecedented precision
in the spectroscopic study of protostars. \textbf{Depending on the orbital period, 
our method allows for the detection of companions within the planetary-mass domain around embedded sources.}\\
\indent We find tentative evidence for the existence of three spectroscopic 
multiple systems out of the total 7 protostars analyzed. With an internal precision in the range of
0.02 to 1.22 kms$^{-1}$ we detected clear RV variations on the order of 4 to 6 kms$^{-1}$.
When we combined our results with those of other studies, we obtain a 
binary fraction of $\simeq$71.5\% for these systems, which 
{is in line with the idea that multiplicity is high at protostellar ages.
The existence of such young SBs strengthens the notion that dynamical evolution has already taken place in the Class I/ II phase (ages $\sim10^{5}$ yrs). \\
\indent Future observations are needed to enhance the sample statistics and will aim to derive the orbital solutions 
for these SBs if their multiple system status is confirmed. \\
\begin{acknowledgements}
We acknowledge the support by the European Research Council/European Community under the FP7 through Starting Grant agreement number 
239953. NCS also acknowledges the support from Funda\c{c}\~ao para a Ci\^encia e a Tecnologia (FCT) through program Ci\^encia\,2007 
funded by FCT/MCTES (Portugal) and POPH/FSE (EC), and in the form of grant reference PTDC/CTE-AST/098528/2008. PVA 
also acknowledges the support from Funda\c{c}\~ao para a Ci\^encia e a Tecnologia (FCT) through program Ci\^encia\,2007 
funded by FCT/MCTES (Portugal) and POPH/QREN, and in the form of grant reference SFRH/BD/47537/2008. PF was supported by the European Research Council/European 
Community under the FP7 through Starting Grant agreement number 239953, as well as by Funda\c{c}\~ao para a Ci\^encia e a Tecnologia (FCT) 
in the form of grant reference PTDC/CTE-AST/098528/2008. We would also like to thank the referee for the effort to critically review 
this manuscript, which has lead to its substantial improvement.

\end{acknowledgements}
\bibliography{maria}
\bibliographystyle{aa}
\addtolength{\tabcolsep}{-3pt}
\longtab{1}{
\begin{longtable}{ >{\tiny}c>{\tiny}c>{\tiny}c>{\tiny}c>{\tiny}c>{\tiny}c>{\tiny}c>{\tiny}c>{\tiny}c>{\tiny}c>{\tiny}c>{\tiny}c }
\caption{From left to right we present the star-forming region of each target, the coordinates, 
observation date, integration time of our observations, K-magnitude,
 the number of nodding cycles observed, the S/N*, the signal to noise ratio of
 the extracted spectra before telluric removal, the 
spectral energy distribution (SED), the reference from where the 
SEDs were collected and an alternative name for each source, if available.}\label{table:1}\\
\hline\hline                 % inserts double horizontal lines
\centering                     % used for centering table
 Source & Region & $\alpha(2000)$ & $\delta(2000)$ & Date & DIT(sec) & K (mag)  & nod.cy & S/N* & SED & reference &alias \\
\hline
\endfirsthead
\caption{Cont.}\\
\hline\hline
\centering 
 Source & Region & $\alpha(2000)$ & $\delta(2000)$ & Date & DIT (sec)  & K(mag)  & nod.cy & S/N* & SED & reference & alias\\
\hline
\endhead
\hline
\endfoot
\hline
\endlastfoot\vspace*{0.001cm}\\
Ced 110 IRS6 & Cha & 11 07 09.80 & -77 23 04.4 & 2008-04-28 & 180.0 & 10.9 &  2 & 25 & FS & Hai06 &  PCW91  \\
&&&& 2008-12-27 & 180.0 & &  2 & 30 &&\\
&&&& 2008-12-28 & 180.0 & &  2 & 30 &&\\
&&&& 2009-02-23 & 180.0 & &  2 & 25 && \\
Cha I T29    & Cha & 11 07 59.25 & -77 38 43.9 &  2008-04-28 & 30.0 & 7.2 &  2 & 25 & Class 0 & Hai04 , Hai06 & V* FK Cha \\
&&&& 2008-12-27 & 30.0 & &  2 & 30 &&\\
&&&& 2008-12-28 & 30.0 & &  2 & 20 &&\\
&&&& 2009-02-23 & 30.0 & &  2 & 20 &&\\
ISO-Cha I 26 & Cha & 11 08 04.00 & -77 38 42.0 & 2008-04-28 & 45.0 & 8.2 &  2 & 30 & Class I & Hai04 , Hai06 & HD 97048 2 \\
&&&& 2008-12-28 & 45.0 & &  2 & 25 &&\\
&&&& 2009-02-23 & 45.0 & &  2 & 30 &&\\
Cha I T32  & Cha & 11 08 04.61 & -77 39 16.9 & 2008-04-28 & 30.0 & 6.1 &  2 & 25 & Class 0 & Hai04 , Hai06 &  \\
&&&& 2008-12-28 & 30.0 & &  2 & 25 &&\\
&&&& 2009-02-23 & 30.0 & &  2 & 25 &&\\
Cha I T33B & Cha & 11 08 15.69 & -77 33 47.1 & 2008-04-28 & 30.0 & 6.9 &  2 & 30 & FS & Hai06 & Glass Ia \\
&&&& 2008-12-28 & 30.0 & &  2 & 35 &&\\
&&&& 2009-02-23 & 30.0 & &  2 & 30 &&\\
Cha I C9-2 & Cha & 11 08 37.37 & -77 43 53.5 & 2008-04-28 & 45.0 & 8.6 &  2 & 20 & Class 0 & Hai04 , Hai06 &  \\
&&&& 2008-12-28 & 45.0 & &  2 & 20 &&\\
&&&& 2009-02-10 & 45.0 & &  2 & 25 &&\\
&&&& 2009-02-26 & 45.0 & &  2 & 20 &&\\
Cha I C1-6 & Cha & 11 09 23.30 & -76 34 36.2 &  2008-04-28 & 45.0 & 8.4 &  2 & 25 & Class I & Hai04 , Hai06 & CCE98 1-76 \\
&&&& 2008-12-28 & 45.0 & &  2 & 30 &&\\
&&&& 2009-02-26 & 45.0 & &  2 & 25 &&\\
Cha I T41  & Cha & 11 09 50.39 & -76 36 47.6 & 2008-04-28 & 30.0 & 7.0 &  2 & 30 & - & - &  \\
&&&& 2008-12-28 & 30.0 & &  2 & 25 &&\\
&&&& 2009-02-26 & 30.0 & &  2 & 25 &&\\
Cha I T42  & Cha & 11 09 54.66 & -76 34 23.7 & 2008-04-28 & 30.0 & 7.0 &  2 & 20 & Class 0 & Hai04 , Hai06 & V* FM Cha \\
&&&& 2009-02-10 & 30.0 & &  2 & 20 &&\\
&&&& 2009-02-26 & 30.0 & &  2 & 20 &&\\
Cha I T44  & Cha & 11 10 01.35 & -76 34 55.8 & 2008-04-28 & 30.0 & 6.4 &  2 & 20 & Class 0 & Hai04 , Hai06 & CHXR 44 \\
&&&& 2009-02-10 & 30.0 & &  2 & 30 &&\\
&&&& 2009-02-26 & 30.0 & &  2 & 25 &&\\
Cha I T47  & Cha & 11 10 50.78 & -77 17 50.6 & 2008-04-28 & 45.0 & 8.8 &  2 & 20 & - & - & V* FO Cha \\
&&&& 2009-02-11 & 45.0 & &  2 & 30 &&\\
&&&& 2009-02-26 & 45.0 & &  2 & 20 &&\\
Cha II 8   & Cha & 12 53 42.88 & -77 15 05.7 & 2008-05-11 & 45.0 & 8.8 &  2 & 25 & - & - &  \\
&&&& 2008-08-17 & 45.0 & &  2 & 20 &&\\
&&&& 2008-08-21 & 45.0 & &  2 & 30 &&\\
GSS 26     & $\rho$Oph & 16 26 10.28&-24 20 56.6 & 2008-04-28 & 120.0 & 9.4 &  2 & 25 & Class I & Bar05 & \\
&&&& 2008-05-09 & 120.0 & &  2 & 35 &&\\
&&&& 2008-08-21 & 120.0 & &  2 & 30 &&\\
GSS/IRS 1  & $\rho$Oph & 16 26 21.50&-24 23 07.0 & 2008-04-28 & 120.0 & 9.0 &  2 & 30 & Class I & Bar05  \\
&&&& 2008-05-09 & 120.0 & &  2 & 20 &&\\
&&&& 2008-09-01 & 120.0 & &  2 & 30 &&\\
GY 23	   & $\rho$Oph & 16 26 24.00&-24 24 49.9 & 2008-04-28 & 30.0 & 7.4 &  2 & 30 & Class II & Bar05 & S2 \\
&&&& 2008-05-09 & 30.0 & &  2 & 25 &&\\
&&&& 2008-09-01 & 30.0 & &  2 & 30 &&\\
GY 51	   & $\rho$Oph & 16 26 30.49&-24 22 59.0 & 2008-04-28 & 180.0 & 10.2 &  2 & 30 & FS & Hai06 & VSSG27 \\
&&&& 2008-05-09 & 180.0 & &  2 & 30 &&\\
&&&& 2008-05-11 & 180.0 & &  2 & 35 &&\\
&&&& 2008-09-01 & 180.0 & &  2 & 25 &&\\
WL 12	   & $\rho$Oph & 16 26 44.30&-24 34 47.5 & 2008-04-28 & 180.0 & 10.4 &  2 & 15 & Class I & Bar05 & GY111 \\
&&&& 2008-05-09 & 180.0 & &  2 & 20 &&\\
&&&& 2008-09-02 & 180.0 & &  2 & 20 &&\\
WL 1 S     & $\rho$Oph & 16 27 04.13&-24 28 30.7 & 2008-05-09 & 180.0 & 10.8 &  2 & 25 & Class II & Bar05  \\
&&&& 2008-05-13 & 180.0 & &  2 & 20 &&\\
WL 17	   & $\rho$Oph & 16 27 06.79&-24 38 14.6 & 2008-05-11 & 180.0 & 10.3 &  2 & 25 & Class I & Bar05 & GY205 \\
&&&& 2008-05-24 & 180.0 & &  2 & 20 &&\\
Elias 29   & $\rho$Oph & 16 27 09.43&-24 37 18.5 &  2008-05-11 & 30.0 & 7.5 &  2 & 25 & Class 0 & Hai04 , Hai06 & GY214 \\
&&&& 2008-05-26 & 30.0 & &  2 & 20 &&\\
&&&& 2008-09-03 & 30.0 & &  2 & 20 &&\\
GY 224     & $\rho$Oph & 16 27 11.17&-24 40 46.7 & 2008-05-11 & 180.0 & 10.8 &  2 & 20 & FS & Bar05 & WLY 1-43 \\
&&&& 2008-05-24 & 180.0 & &  2 & 30 &&\\
&&&& 2008-09-03 & 180.0 & &  2 & 25 &&\\
IRS 34     & $\rho$Oph & 16 27 15.48&-24 26 40.6 & 2008-05-11 & 180.0 & 10.3 &  2 & 25 & FS & Bar05 & GY239 \\
&&&& 2008-05-26 & 180.0 & &  2 & 25 &&\\
&&&& 2008-09-03 & 180.0 & &  2 & 30 &&\\
IRS 37     & $\rho$Oph & 16 27 17.54&-24 28 56.5 & 2008-05-11 & 180.0 & 10.9 &  2 & 25 & Class I & Hai04 & GY244\\
&&&& 2008-05-26 & 180.0 & &  2 & 25 &&\\
&&&& 2008-09-24 & 180.0 & &  2 & 25 &&\\
IRS 42     & $\rho$Oph & 16 27 21.45&-24 41 42.8 & 2008-05-11 & 45.0 & 8.6 &  2 & 25 &  FS & Bar05 & GY252 \\
&&&& 2008-05-24 & 45.0 & &  2 & 30 &&\\
&&&& 2008-09-24 & 45.0 & &  2 & 30 &&\\
WL 6	   & $\rho$Oph & 16 27 21.83&-24 29 53.2 & 2008-05-11 & 180.0 & 10.8 &  2 & 35 & Class I & Bar05  & GY254\\
&&&& 2008-09-24 & 180.0 & &  2 & 30 &&\\
&&&& 2008-09-25 & 180.0 & &  2 & 30 &&\\
IRS 43     & $\rho$Oph & 16 27 26.90&-24 40 51.5 & 2008-05-11 & 120.0 & 9.4 &  2 & 25 & Class I & Bar05 & GY265 \\
&&&& 2008-05-24 & 120.0 & &  2 & 20 &&\\
&&&& 2008-09-03 & 120.0 & &  2 & 25 &&\\
IRS 44     & $\rho$Oph & 16 27 28.00&-24 39 34.3 & 2008-05-11 & 120.0 & 9.7 &  2 & 30 & Class I & Bar05 & GY269\\
&&&& 2008-05-24 & 120.0 & &  2 & 20 &&\\
&&&& 2008-09-03 & 120.0 & &  2 & 30 &&\\
VSSG 18    & $\rho$Oph & 16 27 28.44&-24 27 21.9 & 2008-05-11 & 120.0 & 9.2 &  2 & 30 & FS & Bar05  \\
&&&& 2008-05-26 & 120.0 & &  2 & 30 &&\\
&&&& 2008-09-24 & 120.0 & &  2 & 30 &&\\
IRS 46     & $\rho$Oph & 16 27 29.70&-24 39 16.0 & 2008-05-11 & 180.0 & 10.6 &  2 & 20 & Class I & Bar05 & GY274 \\
&&&& 2008-05-26 & 180.0 & &  2 & 20 &&\\
&&&& 2009-02-23 & 180.0 & &  2 & 30 &&\\
GY 279     & $\rho$Oph & 16 27 30.18&-24 27 44.3 & 2008-05-11 & 45.0 & 9.0 &  2 & 30 & FS & Bar05 & IRS47\\
&&&& 2008-05-26 & 45.0 & &  2 & 20 &&\\
&&&& 2008-09-25 & 45.0 & &  2 & 25 &&\\
IRS 48     & $\rho$Oph & 16 27 37.20&-24 30 34.0 &  2008-05-11 & 30.0 & 7.7 &  2 & 20 & FS & Hai06 & GY304 \\
&&&& 2008-05-26 & 30.0 & &  2 & 25 &&\\
&&&& 2009-02-22 & 30.0 & &  2 & 25 &&\\
IRS 51 S   & $\rho$Oph & 16 27 39.84&-24 43 16.1 & 2008-08-21 & 45.0 & 8.7 &  2 & 20 & FS & Hai06 & GY315\\
&&&& 2008-09-03 & 45.0 & &  2 & 25 &&\\
&&&& 2009-02-22 & 45.0 & &  2 & 30 &&\\
IRS 54     & $\rho$Oph & 16 27 51.70&-24 31 46.0 & 2008-08-21 & 180.0 & 10.2 &  2 & 20 & Class I & Hai06 & GY378 \\
&&&& 2009-02-12 & 180.0 & &  2 & 25 &&\\
&&&& 2009-02-22 & 180.0 & &  2 & 30 &&\\
IRS 63     & $\rho$Oph & 16 31 35.53&-24 01 28.3 &  2008-08-21 & 120.0 & 9.3 &  2 & 25 & Class I & Hai02 , Hai06 & WLY 2-63\\
&&&& 2009-02-11 & 30.0 & &  2 & 30 &&\\
&&&& 2009-02-23 & 30.0 & &  2 & 20 &&\\
L1689SNO2N & $\rho$Oph & 16 31 52.13&-24 56 15.2 & 2008-08-21 & 45.0 & 8.3 &  2 & 30 & FS & Hai06 \\
&&&& 2009-02-11 & 45.0 & &  2 & 20 &&\\
&&&& 2009-02-22 & 45.0 & &  2 & 25 &&\\
IRS 67     & $\rho$Oph & 16 32 01.00&-24 56 44.0 & 2008-08-25 & 180.0 & 10.3 &  2 & 20 & Class I & Hai02 , Hai06 \\
&&&& 2009-02-11 & 180.0 & &  2 & 30 &&\\
&&&& 2009-02-23 & 180.0 & &  2 & 20 &&\\
SVS 20 S   & Serpens & 18 29 57.70 & +01 14 07.0 & 2008-05-09 & 30.0 & 7.1 &  2 & 20 & FS & Hai06 \\
&&&& 2008-05-22 & 30.0 & &  2 & 20 &&\\
&&&& 2008-05-26 & 30.0 & &  2 & 25 &&\\
&&&& 2008-08-21 & 30.0 & &  2 & 30 &&\\
EC 95	   & Serpens & 18 29 57.80 & +01 12 52.0 & 2008-05-09 & 120.0 & 9.8 &  2 & 25 & Class II & Hai06\\
&&&& 2008-05-22 & 120.0 & &  2 & 20 &&\\
&&&& 2008-05-26 & 120.0 & &  2 & 30 &&\\
&&&& 2008-08-21 & 120.0 & &  2 & 20 &&\\
\hline
\hline
Radial velocity standards&&&&&&&&&&\\
\hline
HD 129642 &  & 14 45 09.74 & -49 54 58.61 & 2008-04-28 & 120.0 & 6.2 &  3 & 110 &  &   \\
&&&& 2008-05-09 & 120.0 & &  3 & 120 &&\\
&&&& 18 aug 2008 &120.0 & &  3 & 130 &&\\
&&&& 2008-09-03 & 120.0 & &  3 & 120 &&\\
HD 105671 &  & 12 09 54.98& -46 12 30.20 & 2008-12-28 & 60.0 & 5.8 &  2 & 150 &  &   \\
Gl 406 &  & 10 56 28.86 & +07 00 52.77  & 28 feb 2009 & 60.0 & 6.1 &  2 & 120 &  &   \\
\hline %inserts single line
\end{longtable}
}
% End \longtab 
\end{document}